# The generation of promoter-mediated transcriptional noise in bacteria

Namiko Mitarai[1], Ian B. Dodd [2], Michael T. Crooks[2], and Kim Sneppen[3]*,

[1] Department of Physics, Kyushu University 33, Fukuoka 812-8581, Japan.

[2] Discipline of Biochemistry School of Molecular and Biomedical Science, University of Adelaide, SA 5005, Australia.

[3] Niels Bohr Institute, Blegdamsvej 17, DK-2100, Copenhagen, Denmark.

**Abstract:** Noise in the expression of a gene produces fluctuations in the concentration of the gene product. These fluctuations can interfere with optimal function or can be exploited to generate beneficial diversity between cells; gene expression noise is therefore expected to be subject to evolutionary pressure. Shifts between modes of high and low rates of transcription initiation at a promoter appear to contribute to this noise both in eukaryotes and prokaryotes. However, models invoked for eukaryotic promoter noise such as stable activation scaffolds or persistent nucleosome alterations seem unlikely to apply to prokaryotic promoters. We consider the relative importance of the steps required for transcription initiation. The 3-step transcription initiation model of McClure is extended into a mathematical model that can be used to predict consequences of additional promoter properties. We show in principle that the transcriptional bursting observed at an *E. coli* promoter by Golding *et al.* (2005) can be explained by stimulation of initiation by the negative supercoiling behind a transcribing RNA polymerase (RNAP) or by the formation of moribund or dead-end RNAP-promoter complexes. Both mechanisms are tunable by the alteration of promoter kinetics and therefore allow the optimization of promoter mediated noise.

**Author Summary:** Noise in gene expression is important for phenotypic variation among genetically identical cells. The gene expression will be particularly sensitive to noise in transcription initiation. Transcription initiation from a given promoter involves multiple steps, each of which could be rate limiting. In this paper we discuss how transcription initiation could come in bursts, separated by long periods where the promoter is inactive. Our results are compared to recent data of Golding *et al.* (2005) which suggest that transcriptions from some prokaryotic promoters occur in a highly irregular burst-like fashion. We show that the observed bursting could be caused by one of two alternate mechanisms. One possibility is that changes in supercoiling induced by previous RNA polymerase can help a subsequent RNAP to enter directly into open complex. Another possibility is that an RNAP at the promoter sometimes form a dead-end complex, and thereby occlude the promoter for a sizeable amount of time.

---

* Corresponding author. E-mail: sneppen@nbi.dk

# Introduction

Cellular processes involve stochastic reactions between limited numbers of molecules, and therefore are subject to random noise. The existence of noise in the intracellular concentration of various species has been highlighted in a number of natural and engineered genetic circuits [1-6], which has been coupled with an increasing focus on the theory of how noise might be controlled or exploited by the cell.

Gene expression is perhaps the most important stochastic process in the cell. Transcription involves the production of small numbers of mRNAs, which are then translated multiple times, creating and amplifying noise in protein concentrations. Therefore, the probability distribution underlying the timing of transcription initiation is important for understanding cellular dynamics. A distribution where initiations are evenly spaced will result in less noise and a more uniform cell population. In contrast, a highly variable rate of initiation will produce large fluctuations that can lead to heterogeneous behavior across populations of genetically identical cells. This variability is important to allow populations of unicellular organisms to cope with variable environments [1, 5]. Another example is the spontaneous induction of 'non-inducible' prophages such as P2 [7], where stochastic flipping of a genetic switch allows a low rate of transition from lysogeny into lytic development. Noise in transcriptional initiation also has implications for transcriptional interference between convergent promoters [8]. Bertrand [9] and colleagues have developed a system where an mRNA containing multiple MS2 binding sites can be visualized by the binding of MS2-GFP fusion proteins to the mRNA. Golding and colleagues [10] placed such an mRNA under the control of the $P_{lac/ara}$ promoter in *E. coli* and could thereby detect production of individual mRNAs. When the promoter was induced, transcription was observed to occur in an unexpectedly irregular fashion, with bursts of transcription separated by long periods of inactivity. This phenomenon was called transcriptional bursting. The bursts of activity (on-periods) lasted an exponentially distributed amount of time, with a mean of 6 minutes at 22°C. During an on period a geometrically distributed number of transcripts are produced in rapid succession, with a mean of 2.2 transcripts per on-period. The long periods without transcription (off-periods) were also exponentially distributed, with a mean of 37 minutes. Golding *et al.* also report that similar behavior is seen with the P$_{RM}$ promoter of phage lambda.

Golding *et al.* [10] showed that this behavior was inconsistent with transcription occurring as a Poisson process. Here we consider the McClure model of transcription initiation [11-13], a more general model of transcription initiation, and show that it is still unable to reproduce the transcriptional bursting observed by Golding *et al.* We then consider current hypotheses for the mechanism of transcriptional bursting and find them wanting. Finally we propose two novel hypotheses for the mechanism behind transcriptional bursting, demonstrating that they are able to explain the results of Golding *et al.*

## Results

### *The standard promoter model does not produce bursting*

Golding *et al.* showed that their results were not consistent with transcription initiation being a single Poisson process. By considering the McClure model of transcription initiation (figure 1A) we show that initiation as a single Poisson process is a special case where only one step is rate limiting, and that while the more general case is not a single Poisson process it is still unable to fit the results of Golding *et al.*

In prokaryotes, the initiation of transcription requires the binding of an RNAP to the promoter, the isomerisation of the RNAP through several intermediate forms, rounds of abortive initiation and then finally release from the promoter. Here we consider the McClure model of transcription [11-13] (figure 1(A)), where transcription initiation requires three steps: RNA polymerase (RNAP) binding to the promoter to form a closed complex, followed by isomerisation of the closed complex to an open complex in which the DNA at the promoter is melted, and the escape of the open complex to form an RNAP complex engaged in elongation of the transcript. The closed complex is assumed to be in rapid equilibrium with free RNAP, while isomerisation and escape are treated as being slower and irreversible. This model is a simplified but useful version of the full kinetics of initiation.

The kinetics of each elementary reaction in initiation determines the final distribution of transcription initiation. Transcription is often treated as a Poisson process, i.e. the probability of initiation at a given moment is a constant, which results in an exponential distribution of times between transcripts. Golding *et al.* were able to show through several methods that the distribution of transcription initiation was non-Poisson. However, the exponential distribution is a special case where there is only one rate limiting step in the initiation of transcription.

For the analytical analysis of the McClure model, we make the assumption that the rates of binding $k_b$ and unbinding $k_u$ of the closed complex are relatively fast, and therefore that there are only two kinetically significant steps, isomerisation of the closed complex to an open complex, and promoter escape by the open complex. We assume that each step is elementary, i.e. that it can be approximated as a single chemical reaction. We also ignore the effect of self-occlusion, where an RNAP prevents further initiation at the promoter until it has transcribed far enough to no longer occlude the promoter (50bp), as the time needed to transcribe this distance (1-4 seconds) is negligible compared to the time between initiations in the Golding *et al.* experiments. The average time needed to complete the first step, $\tau_o$, is therefore $\tau_o=(1+K)/O$, where $K=k_u/k_b$ is the equilibrium constant of dissociation for the closed complex and $O$ is the rate of transition from closed to open complex. The inverse of the rate of the open to elongating transition ($E$) gives the average time needed for the second kinetically important step, $\tau_E$ (Fig. 1a). The average time taken for initiation (and therefore the time gap between initiations, $\langle \Delta t \rangle$, with $\langle \cdots \rangle$ indicating the average) is the sum of two exponentially distributed random variables, $\langle \Delta t \rangle = \langle \tau_O + \tau_E \rangle$. The probability distribution of time gaps between initiations is given by

$$P(\Delta t) = \frac{\left[\exp\left(-\frac{\Delta t}{\tau_O}\right) - \exp\left(-\frac{\Delta t}{\tau_E}\right)\right]}{\tau_O - \tau_E} \qquad (1)$$

for $\tau_O \neq \tau_E$. For $\tau_O = \tau_E = \tau$, we get

$$P(\Delta t) = \frac{1}{\tau^2} \Delta t \exp\left(-\frac{\Delta t}{\tau}\right). \qquad (2)$$

In the case where one step is much slower than the other (Class I), there is only one rate-limiting step in initiation and the distribution of $\Delta t$ approaches a single exponential with mean $\tau_L=\max(\tau_O,\tau_E)$ (Equation 1; figure 1B), i.e. it approaches a single Poisson process. Here, the data points in figure 1B) have been obtained by simulating the model of the promoter in figure 1A) using the Gillespie algorithm [14], which stochastically determines the next reaction to occur and the time interval between reactions based on the given rates. The other extreme, where $\tau_O = \tau_E$ (Class II), is shown in figure 1C. In Class II, the chance of rapid successive firings faster than the average ($\Delta t \ll \tau_O + \tau_E$) is smaller than for a Class I promoter, as for a Class II promoter a low $\Delta t$ requires both the isomerisation and the escape to productive transcription to occur in rapid succession, whereas for a Class I promoter a low $\Delta t$ requires the rapid occurrence of the rate limiting step only. As a consequence the distribution in Class II shows a peak at non-zero $\Delta t$. Promoter models that specify more kinetically significant reaction intermediates produce more extreme versions of the Class II distribution, with a larger peak centered around $\langle \Delta t \rangle$, resulting in more regular firing intervals.

The Class I type promoter shows the most fluctuation in $\Delta t$, and the effect of adding more kinetically significant intermediate steps is to reduce the amount of variability in $\Delta t$. Therefore neither the standard model nor models that take into account more intermediates can reproduce the bunched activity observed by Golding *et al.* [10], which show greater fluctuations in $\Delta t$ than a Poisson process. In order to reproduce the bunched activity, it is necessary to consider a model with a branched pathway, where the system can go into either an active state or an inactive state with a switching mechanism between them.

*Previously Proposed Mechanisms for Bunched Activity*

Here we consider several hypotheses for the mechanism of transcriptional bursting and argue that they are unlikely to be correct. The promoter used by Golding *et al.* [10], P$_{lac/ara}$, can be repressed about 70 fold by the lac repressor and activated about 30 fold by AraC [15]. Therefore, a simple hypothesis put forward by Golding *et al.* is that the silent periods are periods where the lac repressor is bound to the promoter, and the bursts are periods of activity when the promoter is free. However, the mean duration of off-periods is 37 min while on periods are only 6 min in duration, despite the fact that the promoter has been fully induced by 1mM IPTG. It seems impossible for the lac repressor to remain bound to the DNA for 37 minutes under these conditions; especially considering that 1mM IPTG derepresses the lac promoter in less than 5 sec [16].

A similar idea is that the off-periods represent periods where AraC is not bound to the promoter [10]. To make this feasible the on rate for AraC in an *E.coli* cell would have to be exceedingly small given the large off periods. This is unreasonable in view of the high association rate for AraC to other operators [17]. Presumably association rate is diffusion limited, meaning that it would take one AraC molecule less than a minute to bind to the operator [18]. In conclusion we find it unlikely that binding AraC is sufficient to produce bunched activity.

Another hypothesis put forward by Golding *et al.* is that RNAP might be able to re-initiate after termination, aided by the retention of sigma factor during transcription [19]. Presumably the RNAP would have to be positioned to rebind to the same promoter after termination for re-initiation to occur with any reliability, and it is not clear how this would be caused. One possibility is that a transcription factor might remain in contact with both the RNAP and the promoter via a DNA loop. This would render the promoter unavailable during transcription, which has some support from the data in that the lengths of the observed on-periods were approximately equal to the number of initiations multiplied by the time taken to transcribe the reporter mRNA for both $P_{lac/ara}$ and $P_{RM}$ (Golding, private communication), which would be expected if transcription does not occur simultaneously. However, this data is somewhat anecdotal, and stands in contradiction to the simultaneous transcription observed with electron microscopy [20]. Also, this mechanism requires binding of a closed complex to the DNA to be the rate limiting step that causes the 37 minute long off-period, and we consider it unlikely that simple recognition of the promoter by RNAP would take this long, especially given that closed complex formation is often thought to be a rapid equilibrium process.

Multiple RNAP can cooperate to overcome pause sites [21]. It might therefore be possible that the burst is due to multiple RNAP building up at a pause site and overcoming it together. However, this would require the RNAP to pause for a length of time on the same scale as the off-period; such an extreme pause is unlikely given that even the strongest pauses measured *in vitro* only last for around one minute.

Bursting could also result if there were distinct regions of high and low transcriptional activity within bacteria, akin to the idea of transcription factories in eukaryotes, and the promoter moved in and out of these regions on a slow time scale [22, 23]. Although this is an interesting possibility, not enough is known to evaluate such a mechanism in bacteria in much detail.

Fluctuations in the availability of free RNAP within the cell could contribute to variable initiation rates but it is difficult to see how such severe and long-lasting fluctuations capable of producing extended periods of complete inactivity could occur in cells where ~3000 RNAPs [24] produce >$10^5$ RNAs per generation.

*Super-coiling mediated recruitment*

There is both theoretical [25] and experimental evidence [26, 27] that an elongating

RNAP can increase the negative supercoiling of the DNA behind it.

Promoters can be very sensitive to supercoiling; for example, *in vitro* the activity of the LacP promoter increases by more than a factor of 10 when the super-coiling σ is changed from zero to -0.065 (which is the average supercoiling of DNA in E. coli) [26]. We therefore consider it a possibility that the bursts of transcription might be caused by a transcribing RNAP assisting the recruitment of further RNAP via the wake of supercoiling left behind it. In principle one could argue that perturbed supercoil states could relax quickly in a plasmid [25] like the one used by Golding *et al.*, but it has been demonstrated that a promoter can induce huge changes in supercoiling of a plasmid [28].

Consider a promoter where open complex formation is a rate limiting step that is assisted by negative supercoiling. To model this, we assume that the negative supercoiling assists this step to the extent that it is no longer rate limiting. We parameterize this effect of supercoiling into a single number $q$, the probability that supercoiling left in the wake of a prior RNAP allows a subsequent RNAP to rapidly form an open complex before the supercoiling is relaxed (figure 2A). This then creates two possible behaviors at the promoter. If the promoter is in the supercoiled state, open complex formation is enhanced to the point where it is not rate limiting, and transcription events occur at rate $E$ and are exponentially distributed. If the promoter is not in the supercoiled state, then open complex formation is very much slower and now rate limiting; transcriptional events are still exponentially distributed but now with the much lower rate $O$. This creates the long periods of inactivity associated with off periods (figure 3A) and holds when $O \ll E$, and gives a distribution

$$P(\Delta t) = \frac{q}{\tau_E}\exp\left(-\frac{\Delta t}{\tau_E}\right) + \frac{(1-q)}{\tau_O}\exp\left(-\frac{\Delta t}{\tau_O}\right) \qquad (3)$$

(shown in figure 3B).

The supercoiling need not persist for the full length of the on-period, or for the length of time between two initiations. In the scheme we present here, it is only required that the supercoiling persists long enough to allow an open complex to form rapidly. The final escape step is assumed to be neutral with respect to supercoiling and hence as soon as an open complex has formed at the promoter the supercoiling can be relaxed without interrupting the on-period. This assumption can be varied without changing the general behavior of the model.

If the supercoiling is relaxed before an open complex is formed, the promoter has switched to an off-period where initiation occurs at a much slower rate. The parameter $q$ determines the size of the on-periods, as after each initiation there is a probability $q$ that another open complex will be recruited and the on-period will continue, or a probability 1-$q$ that an off-period will start. Therefore, the probability of getting a burst of $\langle \Delta n \rangle$ initiations is proportional to $q^{\Delta n-1}$. In this model a promoter is in the on-state when it is in the supercoiled state or when it has an open complex. Table 1 gives equations relating model parameters to the average $\langle \Delta n \rangle$, $\langle t_{on} \rangle$, and $\langle t_{off} \rangle$ (Derivations are given in Text S1).

This mechanism can reproduce the observations of Golding *et al.* [10] with the parameters $\tau_O$=37 [min], $\tau_E$=29 [min] and $q$ =0.545. We simulated the recruitment model using the Gillespie algorithm [14]. It gives the expected shape for the $P(\Delta t)$ distribution (figure 3B) and matches the distribution of $\Delta n$ measured by Golding *et al.* (3C) and also the distributions of on and off-periods measured by Golding *et al.* (3D). In these plots the on-periods are defined as being the time intervals when there is rapid successive initiation (figure 3A), following the procedure in Golding *et al.* [10]; the detailed definition is given in the Materials and Methods section.

## Formation of a dead-end complex

Another possibility is that the off periods are due to the formation of long-lived non-productive initiation complexes at the promoter [29-31]. These non-productive complexes have been observed *in vitro* and may be arrested backtracked complexes or complexes that cannot exit the abortive initiation state into productive elongation. In both cases initiation can be made more efficient by the GreA/B RNAP-binding factors [29, 30]. The random formation of such 'dead-end' complexes could block the promoter for extended periods of time, causing productive transcription to be confined to those times when the promoter is free. For the promoter λPR the lifetime of these complexes was found to be in the order of 10-20 minutes under *in-vitro* conditions, thus dead-end complexes can last long enough to cause the observed off-periods [31].

For the analytical treatment of this model we call the probability that a promoter bound complex will undergo a productive initiation $Q$, and the probability that the promoter bound complex enters a moribund state is therefore 1- $Q$. We assume that removal of the moribund complexes is a Poisson process with a rate $d$, which gives $\langle t_{off} \rangle = \tau_{dead}/Q$ with $\tau_{dead}$=1/$d$, which allows for the fact that a single off-period can be caused by multiple subsequent moribund complexes (Table 1). Here we consider a promoter to be in the off-period if it is occupied by dead-end complexes; otherwise it is on. The derivations of on- and off-times are given in Text S1. The dead-end complex mechanism is also capable of causing the behavior observed by Golding *et al.*. The data of Golding *et al.* are reproduced with $Q$=0.545, $\tau_{dead}$=20 [min], and $\tau_O+\tau_E$=2.9 [min]. Figure 3E shows the distribution $P(\Delta t)$ with these parameters obtained by the simulation using the Gillespie algorithm [14]. It has been confirmed that the distributions of $\Delta n$, $t_{on}$, and $t_{off}$ are reproduced as well as the recruitment model (data not shown).

The formation of dead-end complexes is favored by low temperatures at the lac UV5 promoter [32]. If this were also the case for the $P_{lac/ara}$ promoter, it could be part of the explanation for why the $P_{lac/ara}$ promoter is so weak in the conditions used by Golding *et al.* (22°C) when it is reported to be a strong promoter elsewhere [15]. However, the activity of the promoter observed by Golding *et al.* at 37°C is still rather low compared the previously reported estimate [15]. This could be associated with the fact that there is almost no activation of the promoter caused by AraC/arabinose under their experimental conditions (see Fig. 1E in *Golding et al.*). Another possibility could be the presence of an unknown terminator, which would imply that the number of complete transcripts represents only a fraction of the transcription initiation events.

## *Control of transcriptional noise*

One of observations made by Golding *et al.* that was used as evidence for transcriptional bursting was that the Fano factor for the distribution of number of transcripts $N$, $\nu = \langle (N-\langle N \rangle)^2 \rangle / \langle N \rangle$, was approximately 4 for the $P_{lac/ara}$ promoter at 37°C, rather than 1 predicted for Poisson transcription. The Fano factor is a measure of noise; higher values indicating a more noisy process. When the on-periods are much shorter than the off-periods, the Fano factor $\nu$ is linked to the burst size $\Delta n$ as $\nu \approx \langle \Delta n \rangle$. If the on-time is sizable, on the other hand, $\langle \Delta n \rangle$ needs to be much larger to give the same $\nu$.

By considering a population of cells where transcripts are degraded with rate $\gamma$, we can relate $\nu$ to model parameters. Figure 4 shows how $\nu$ varies with model parameters for each model while keeping $\langle N \rangle = 10$ obtained by analytical calculations (The detailed calculations are in the Text S1.). In the recruitment case the Fano factor is larger for smaller $\alpha$ and larger $q$, i.e., when the open complex formation is the rate limiting step and once a firing has occurred further recruitment occurs successively. In the dead-end model the Fano factor is larger for smaller $\beta=(\tau_O+\tau_E)/\tau_{dead}$ and larger $Q$, which occurs when moribund persist for long periods of time, but transcription during the on periods is rapid and occurs many times before another off period occurs. One should note that the Fano factor can be changed depending on parameters for a given $\langle N \rangle$; This means that the noise can be tuned for a given promoter strength under either model, which can allow the promoter noise to evolve to reflect a level that provides the best fitness for the cell.

## Discussion

We have analyzed possible mechanisms of transcriptional bursting in terms of a simple recruitment/isomerisation/escape model. A model where supercoiling created by an RNAP engaged in transcription assists in the recruitment of subsequent RNAPs is able to reproduce all the features of the experiments, without resorting to very large timescales for on-off equilibrium rates, or unknown pause sites or localization effects. Alternatively, the data of Golding *et al.* could also be reproduced if the investigated promoters spent a sizable fraction of their time by being occupied by an RNAP in a non-productive state.

Transcription bursts have been reported in eukaryotic systems [33, 34] and have also been proposed to facilitate cell to cell variability. These eukaryotic model systems both included transcription factors and in addition they may be influenced by chromatin remodeling. The bunched expression of nearby genes is correlated [34], a feature that fits with extended states of chromatin. The dead end complex cannot give such spatial correlations, whereas supercoiling mediated recruitment in principle could correlate expression from two promoters if they are close to each other.

In one mammalian system, the reported pulse duration and silenced periods are similar to the ones modeled in this paper [33]. However, in that system subsequent bursts of transcription are correlated, with one transcription burst priming the system for another one [33], which has not been reported in Golding *et al.*. This is again consistent with the larger scale genomic silencing associated with, for example, chromatin states or the genes repositioning relative to transcriptional factories [22]. The recruitment model cannot account for correlations between subsequent bursts, whereas the dead end model could give such time correlations between busts if the dead end complexes come in different categories, each with their characteristic lifetime.

Overall we stress that our current modeling demonstrates two plausible mechanisms for generating bursts of transcription at an isolated promoter. Additional mechanisms come into play when the promoter is regulated by a transcription factor with a low on-rate, or when large scale reorganization of the chromosome takes place on a slow timescale.

Both the dead-end and the recruitment model can be simulated on-line using the java applet on http://www.cmol.nbi.dk/models/transcription/RNAPInitiation.html.

*Testing the recruitment model*

The recruitment model implies a number of predictions that can be tested. In particular, promoters with bunched transcription initiation will be highly sensitive to negative supercoiling of the DNA. And conversely, promoters that are insensitive to supercoiling will have transcription events which are separated by more regular time intervals.

For promoters that are sensitive to supercoiling, one could selectively shorten the long off periods by introducing a second nearby promoter. One option is to add a divergent promoter that might be able to donate its negative supercoil wake. Such a construct was investigated by Opel *et al.* [27], who reported that a second promoter could indeed increase the activity of a supercoiling sensitive promoter in the *ilvYC* operon. This predicts that if a similar experiment was done with the $P_{lac/ara}$ promoter, then reduced off periods would be observed.

Another prediction is that for promoters with bunched activity the isomerisation step is rate limiting. Thus the fraction of time spent in open complex is small compared to the time between transcription initiations. One might be able to show an inverse correlation between the noisiness of a promoter and the occupancy of the promoter by open complexes using potassium permanganate DNA footprinting [35].

*Testing the dead-end model*

The dead-end mechanism implies that the promoter is mostly occluded by an RNAP with an open transcription bubble. This could be identified permanganate footprinting [35].

The availability of GreA/B could affect the rate of removal of the dead-end complex, $d$ [29, 30]. Overexpression of GreA/B could increase $d$ and reduce off-periods, while longer off-periods, due to lower $d$, could be observed in greA/B mutants.

It is possible that the dead-end complexes could be removed by a collision with an RNAP transcribing from a second promoter in a fashion similar to the removal of an open complex by transcriptional interference [36]. The off-times of a promoter could therefore in principle be shortened by using other RNAP's initiated from another promoter that transcribes across the promoter in question. If a promoter spent a substantial fraction of the time occupied by a dead-end complex, it could be strongly activated by tandem or even convergent promoters, which would be a novel twist on the usually repressive effect of transcriptional interference. If $d$ is reduced in Table 1, the "off-times" could be reduced by a factor set by the ratio of the strength of the two promoters, and the promoter activity could increase. Thus, if $P_{lac/ara}$ activity is affected by dead-end complex formation, then placing a weak divergent promoter upstream should not increase $P_{lac/ara}$ activity but placing this promoter in a convergent orientation may activate $P_{lac/ara}$.

## *Perspectives for the regulation of transcriptional noise*

The sensitivity of a promoter to supercoiling mediated recruitment or dead-end complex formation provides additional avenues for control of overall promoter strength, either by evolution or by regulatory factors.

DNA supercoiling can increase or decrease promoter activity both *in vitro* [26] and *in vivo* [37] in a promoter specific manner. Supercoiling can affect RNAP binding to the promoter and open complex formation *in vitro* and presumably can affect other steps as well. RNAP recruitment induced by the supercoiling created by an elongating transcription complex may contribute significantly to the activity of certain promoters. We expect that, except for very active promoters, rapid dissipation of the supercoil wake would make inhibition of a supercoiling-repressed promoter by this mechanism unlikely. Stimulation by the departing elongating complex should similarly only apply to the early steps in initiation. Thus only promoters whose early steps are rate-limiting and can be enhanced by supercoiling should be stimulated by this mechanism.

The reduction of promoter activity by the formation of dead-end complexes is potentially very strong. The effect increases with the probability of forming such a complex (1-$Q$) and with the lifetime of the complex (1/$d$), parameters which could be determined both by the promoter sequence and by the availability of factors such as GreA/B that may remove the complex [29, 30]. This mechanism would seem to be an inefficient way to set the strength of a promoter, as it would sequester an RNAP. However, it would allow regulation by transcription factors that change the fraction of RNAPs that enter into dead-end complexes or that stabilized the dead-end complex. As a consequence, genes which are silenced through this mechanism will have relatively high fluctuations in expression level, and thereby some cells can explore advantages afforded by relatively high expressions, even when most cells are kept at near zero expression. Bunched activity for a near silenced promoter could, for example, be important in the pathway for the spontaneous induction of lysogeny for some temperate phages, like P2.

High noise in protein levels can also be obtained at the translation level. If a single mRNA molecule is rapidly translated many times the result is a burst of protein production. Therefore transcriptional bursting is not strictly required for protein production to occur in bursts. However, transcriptional bursting might allow for additional modes of regulation by transcription factors or other proteins that influence the state of the DNA around the promoter site. It may also complement bursts of protein production produced by rapid translation by removing constraints placed on burst size by the upper limits of mRNA translation rate.

Dynamics and the interplay between timescales presents an open, and until recently, quite unexplored part of molecular biology. The present analysis suggests a new mechanism for *in vivo* regulation, where long silent timescales emerge as the result of some particularly large rate limiting step in the promoter. These steps are open for new levels of regulation by transcription factors, which naturally will be most effective when they influence the rate limiting step of transcription initiation [38].

## Materials and Methods

### *Calculation methods*

To calculate the activity of a promoter we first calculate the probability that the promoter will be occupied by closed ($\eta$) and open ($\theta$) complexes using steady state conditions. The total activity of the promoter is given by $F=E\theta$ for the standard model and the recruitment model, and $F=QE\theta$ for the dead-end model. Details of the calculation are found in the Text S1.

The time between subsequent initiations is calculated by considering the time needed for each step as described in the Text S1. For class I there is only one step and the distribution is a simple exponential. For class II there is two steps. If these steps take an average time of $\tau_O$ and $\tau_E$, the total waiting time between events is distributed with

$$P(\Delta t) = \int_0^{\Delta t} \frac{1}{\tau_O} \exp\left(-\frac{s}{\tau_O}\right)$$

$$\times \frac{1}{\tau_E} \exp\left(-\frac{\Delta t - s}{\tau_E}\right) ds, \qquad (4)$$

giving eq.(1) in the main text for $\tau_O \neq \tau_E$. For one $\tau$ much greater than the other, this distribution degenerates into a simple exponential. For $\tau_O = \tau_E$, eq. (4) gives eq.(2) in the main text.

For the recruitment model, the intervals between initiations are partitioned between the supercoiling assisted or unassisted outcomes, with a partitioning ratio given by *q*. Details are in the Text S1. For the dead-end model the distribution is similarly partitioned between the two distributions with a partition ratio given by *Q*. Details are in the Text S1. In the Text S1 we also show how to calculate the distribution of "on" and "off" times from *q* or *Q*. Finally, we calculate the Fano factor $v = \langle (N - \langle N \rangle)^2 \rangle / \langle N \rangle$ by using generating functions as described in the Text S1.

*Protocol to determine on-periods and off-periods*

We distinguish "on-periods" and "off-periods" in the simulation data following the procedure used by Golding *et al.* [10]. They analyzed the experimentally obtained time series of fluorescent signal manually. The system is considered to be in "off-period" when the signal does not change for a while, and otherwise it is in "on-period". The specific time resolution to detect an "off-period" was not given, but the shortest off-time measured was around 6 [min] (Golding, private communication); in other words, transcription events separated by less than 6 [min] were considered to be in the same "on-period".

During an on-period, the number of messages transcribed, $\Delta n \geq 1$, and the duration $t_{on}$ were recorded; the time to transcribe one message $\Delta$ was 2.5 [min] [10], which corresponds to the on-time for $\Delta n=1$ case.

Considering this protocol used by Golding *et al.* [10], we defined $\Delta n$, $t_{on}$, and the duration of the off-time $t_{off}$ out of the time series of firings from our model (figure 3A) as follows: (i) When firings are separated by more than $\tau_c$=6 [min]+$\Delta$=8.5[min], the promoter is in an off period. (iii) Otherwise, if successive firings are separated by an interval less than $\tau_c$, the gene is considered to be on until we observe an interval greater than $\tau_c$. This defines the on-time $t_{on}$, and we count the number of transcripts per on-time $\Delta n$.

**Acknowledgement:** NM and KS thank I. Golding for providing them the detailed data.

# References

1. Suel GM, Kulkarni RP, Dworkin J, Garcia-Ojalvo J, Elowitz MB (2007) Tunability and noise dependence in differentiation dynamics. Science 315: 1716-1719.

2. Thattai M, van Oudenaarden A (2001) Intrinsic noise in gene regulatory networks. Proc Natl Acad Sci U S A 98: 8614-8619.
3. Elowitz MB, Levine AJ, Siggia ED, Swain PS (2002) Stochastic gene expression in a single cell. Science 297: 1183-1186.
4. Raser JM, O'Shea EK (2005) Noise in gene expression: origins, consequences, and control. Science 309: 2010-2013.
5. Blake WJ, KAErn M, Cantor CR, Collins JJ (2003) Noise in eukaryotic gene expression. Nature 422: 633-637.
6. Paulsson J. (2004) Summing up the noise in gene networks. Nature 427: 415-418.
7. Ljungquist E, Bertani LE (1983) Properties and products of the cloned int gene of bacteriophage P2. Mol Gen Genet 192: 87-94.
8. Sneppen K, Dodd IB, Shearwin KE, Palmer AC, Schubert RA, Callen BP, Egan JB (2005) A mathematical model for transcriptional interference by RNA polymerase traffic in Escherichia coli. J Mol Biol. 346: 399-409.
9. Bertrand E, Chartrand P, Schaefer M, Shenoy SM, Singer RH, Long RM (1998) Localization of ASH1 mRNA particles in living yeast. Mol. Cell, 2: 437-445.
10. Golding I, Paulsson J, Zawilski SM, Cox EC (2005) Real-time kinetics of gene activity in individual bacteria. Cell 123: 1025-1036.
11. Hawley DK, McClure WR (1982) Mechanism of activation of transcription initiation from the lambda PRM promoter. J Mol Biol. 157: 493-525.
12. Buc H, McClure WR (1985) Kinetics of open complex formation between Escherichia coli RNA polymerase and the lac UV5 promoter. Evidence for a sequential mechanism involving three steps. Biochemistry, 24: 2712-2723.
13. McClure W.R. (1985) Mechanism and control of transcription initiation in prokaryotes. Annu Rev Biochem. 54: 171-204.
14. Gillespie DT (1977) Exact stochastic simulation of coupled chemical reactions. J. Phys. Chem. 81: 2340-2361.
15. Lutz R, Bujard H (1997), Independent and tight regulation of transcriptional units in Escherichia coli via the LacR/O, the TetR/O and AraC/I1-I2 regulatory elements. Nucleic Acids Res 25: 1203-1210.
16. Elf J, Li GW, Xie XS (2007) Probing transcription factor dynamics at the single-molecule level in a living cell. Science 316: 1191-1194.
17. Timmes A, Rodgers M, Schleif R, (2004) Biochemical and physiological properties of the DNA binding domain of AraC protein. J. Mol. Biol.340: 731-738.
18. Sneppen K, Zocchi G (2005), Physics in Molecular Biology.: Cambridge university press.

19. Bar-Nahum G, Nudler E (2001) Isolation and characterization of sigma(70)-retaining transcription elongation complexes from Escherichia coli. Cell, 106: 443-451.
20. French SL, Miller OL, Jr. (1989) Transcription mapping of the Escherichia coli chromosome by electron microscopy. J Bacteriol. 71: 4207-4216.
21. Epshtein V, Nudler E (2003) Cooperation between RNA polymerase molecules in transcription elongation. Science 300: 801-805.
22. Cook PR (1999) The organization of replication and transcription. Science 284: 1790-1795.
23. Cabrera JE, Jin DJ (2003) The distribution of RNA polymerase in Escherichia coli is dynamic and sensitive to environmental cues. Mol Microbiol, 50: 1493-1505.
24. Pedersen S, Bloch PL, Reeh SV, Neidhardt FC (1978) Patterns of protein synthesis in E. coli: a catalog of the amount of 140 individual proteins at different growth rates. Cell 14: 179-190.
25. Liu LF, Wang JC (1987) Supercoiling of the DNA template during transcription. Proc Natl Acad Sci USA 84: 7024-7027.
26. Lim HM, Lewis DE, Lee HJ, Liu M, Adhya S (2003) Effect of varying the supercoiling of DNA on transcription and its regulation. Biochemistry, 42: 10718-10725.
27. Opel ML, Hatfield GW (2001) DNA supercoiling-dependent transcriptional coupling between the divergently transcribed promoters of the ilvYC operon of Escherichia coli is proportional to promoter strengths and transcript lengths. Mol Microbiol, 39: 191-198.
28. Samul R, Leng F (2007) Transcription-coupled hypernegative supercoiling of plasmid DNA by T7 RNA polymerase in Escherichia coli topoisomerase I-deficient strains. J Mol Biol. 374: 925-935.
29. Stepanova E, Lee J, Ozerova M, Semenova E, Datsenko K, Wanner BL, Severinov K, Borukhov S (2007) Analysis of promoter targets for Escherichia coli transcription elongation factor GreA in vivo and in vitro. J Bacteriol. 189: 8772-8785.
30. Susa M, Kubori T, Shimamoto N (2006) A pathway branching in transcription initiation in Escherichia coli. Mol Microbiol. 59: 1807-1817.
31. Kubori T, Shimamoto N (1996), A branched pathway in the early stage of transcription by Escherichia coli RNA polymerase. J Mol Biol. 256: 449-457.
32. Straney DC, Crothers DM (1985) Intermediates in transcription initiation from the E. coli lac UV5 promoter. Cell 43: 449-459.
33. Chubb JR, Trcek T, Shenoy SM, Singer RH (2006) Transcriptional pulsing of a

developmental gene. Curr Biol, 16: 1018-1025.

34. Raj A, Peskin CS, Tranchina D, Vargas DY, Tyagi S (2006) Stochastic mRNA synthesis in mammalian cells. PLoS Biol. 4: e309.

35. Sasse-Dwight S, Gralla JD (1991) Footprinting protein-DNA complexes in vivo. Methods Enzymol 208: 146-168.

36. Callen BP, Shearwin KE, Egan JB (2004) Transcriptional interference between convergent promoters caused by elongation over the promoter. Mol Cell 14: 647-656.

37. Peter BJ, Arsuaga J, Breier AM, Khodursky AB, Brown PO, Cozzarelli NR (2004) Genomic transcriptional response to loss of chromosomal supercoiling in Escherichia coli. Genome Biol, 2004. 5: R87.

38. Rostoks N, Park S, Choy HE (2000) Reiterative transcription initiation from galP2 promoter of Escherichia coli. Biochim Biophys Acta 1491: 185-195.

Figures:

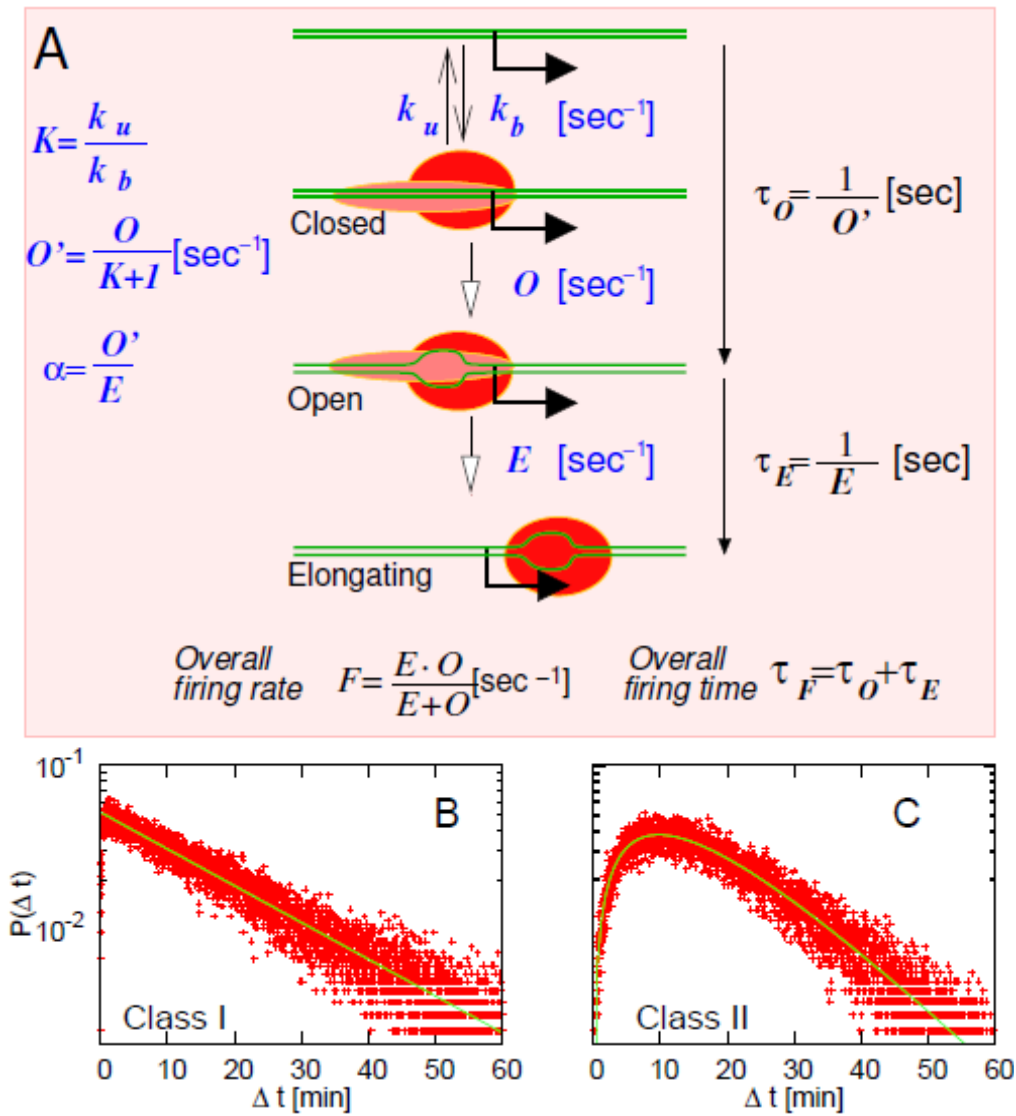

Figure 1. The Hawley-McClure 3-step model of transcription initiation. (A) First an RNAP forms a closed complex at the promoter with some on ($k_b$) and off rates ($k_u$), and subsequently forms an open complex with rate $O$. This process is a directed non-equilibrium transition. Finally, the open complex escapes the promoter into productive transcription with the one way rate $E$. (B) and (C): The distribution of intervals between transcription events for the standard 3-step model. For binding and unbinding rates of the closed complex we use $k_b=k_u=60$ [1/min]. The average total strength is $F=1/(20$ [min]$)$, whereas $l=35$ [bp] and $v=25$ [bp/sec]. (B) Class I: Probability distribution for time between transcripts $\Delta t$ for a promoter with a single rate limiting step. Here, it is isomerisation with $O=1/(9.7$ [min]$)$. The escape rate from the open complex is $E=1/(0.19$[min]$)$. Red dots: stochastic simulation results. Solid line: predicted distribution, $(1/\tau)\exp[-\Delta t/\tau]$ with $\tau=19.4$ [min]. (C) Class II; $\Delta t$ probability distribution for a promoter with two rate limiting steps because the isomerisation and escape rates are similar ($O=1/(4.9$[min]$)$, $E=1/(9.7$

[min])). Solid line shows the predicted distribution, $(1/\tau)^2 \Delta t \exp[-\Delta t/\tau]$ with $\tau$=9.7 [min].

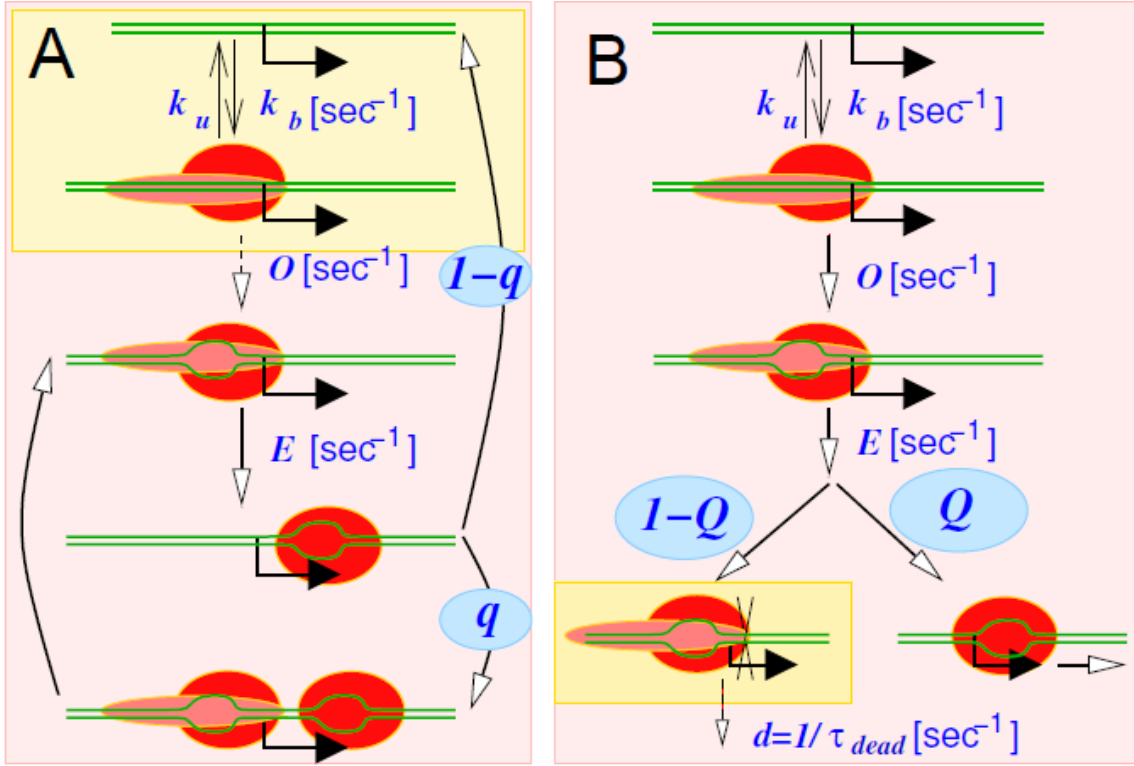

Figure 2. Revisiting mechanisms for bunched firing. We propose (A) supercoiling assisted open complex formation, and (B) possible stalling of an RNAP into a "moribund" complex. The yellow background indicates the state where most of the time is spent in the off period. (A) An elongating RNAP might recruit a subsequent RNAP into an open complex with probability $q$, thus by-passing the time needed to recruit an RNAP and form open complex. In the limit of large $k_b$, the firing rate is given by $F = \frac{E \cdot O'}{O'+(1-q)E} = \frac{O'}{\alpha+(1-q)}$ with $\alpha = \frac{O'}{E}$. (B) Two alternative open complexes, of which one is productive and the other is a dead end complex that is removed with rate $d$. In this case $Q$ denotes the probability that a closed complex enters into the productive open complex. In the limit of large $k_b$, the firing rate is given by $F = \frac{QEO'}{E+O'+(1-Q)O'E/d} = \frac{Qd}{\beta+(1-Q)}$ with $\beta = \frac{\tau_O+\tau_E}{\tau_{dead}} = \frac{d(E+O')}{EO'}$. The detailed calculations and equations for limiting $k_b$ are given in the Text S1. Both the dead-end and the recruitment model can be simulated on-line using the java applet on http://www.cmol.nbi.dk/models/transcription/RNAPInitiation.html.

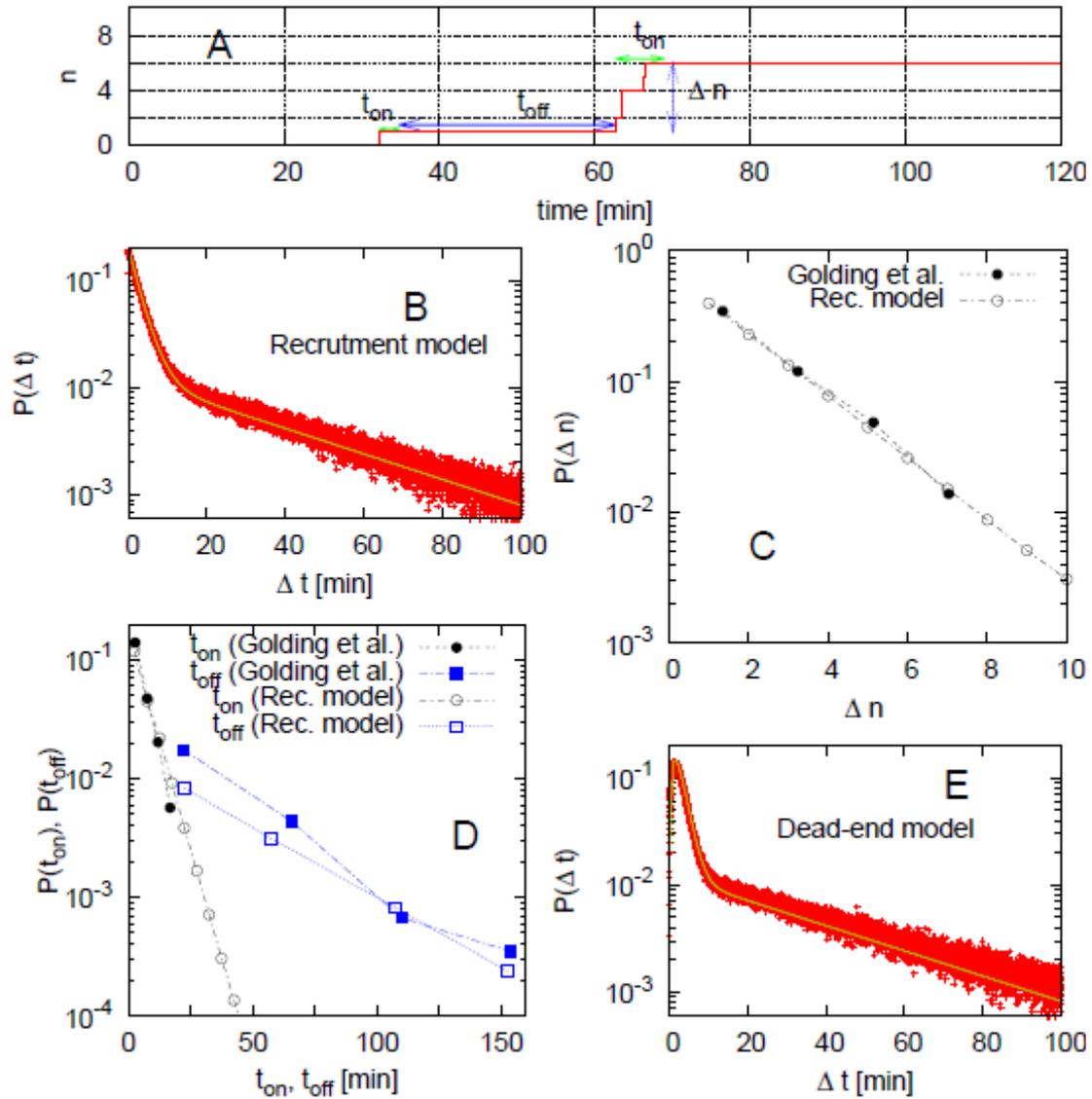

Figure 3. Transcriptional bursting from three step model with supercoiling assisted recruitment and the dead-end complex model. (A) -(D) show the results from the recruitment model with the probability of recruitment $q=0.545$. The average firing rate is $1/20$ [1/min]. $k_b=k_u=60$ [1/min], $E=1/2.9$ [1/min], $O=1/18.3$ [1/min], $l=35$ [bp], $v=25$ [bp/sec]. (A) Accumulated number of mRNAs, showing on periods and off-periods. Transcriptional bursting can be seen around 62 [min] to 70 [min], where 5 firings occur in rapid succession. See Materials and Methods for the choice of parameters and definitions of $t_{on}$ and $t_{off}$. (B) The distribution of the durations between firing, $P(\Delta t)$. The solid line shows eq. **(3)**. (C) The distribution of the number of firings per on-period, $P(\Delta n)$. The filled circles show experimental data from Golding *et al* [10], and the open circles are $\Delta n$ from the recruitment model. (D) The distribution of the "on-times" $t_{on}$ (open circles) and the "off-times" (open boxes) $t_{off}$. The experimental data from Golding *et al.* for the on-time (filled circles) and off-time (filled boxes) are also shown. (E) The distribution of intervals between initiations for the dead-end complex model (figure **2**B). The probability of productive

elongation is $Q=0.545$, and the rate of removal of dead-end complexes is $d=1/\tau_{dead}=1/(20\,[\text{min}])$. The average firing rate is $1/22\,[1/\text{min}]$. $k_b=k_u=60\,[1/\text{min}]$, $E=1/1.5\,[1/\text{min}]$, $O=1/0.7\,[1/\text{min}]$, $l=35\,[\text{bp}]$, $v=25\,[\text{bp/sec}]$. This parameter choice corresponds to a Class 2 promoter in the on periods, which gives a round curve for short timescales (around 3 [min]). The solid line shows

$$Q\Delta t\exp[-\Delta t/\tau]/\tau^2 + (1-Q)\exp[-\Delta t/(\tau_{dead}/Q)]/(\tau_{dead}/Q)\quad\text{with } \tau=\tau_O=\tau_E.$$

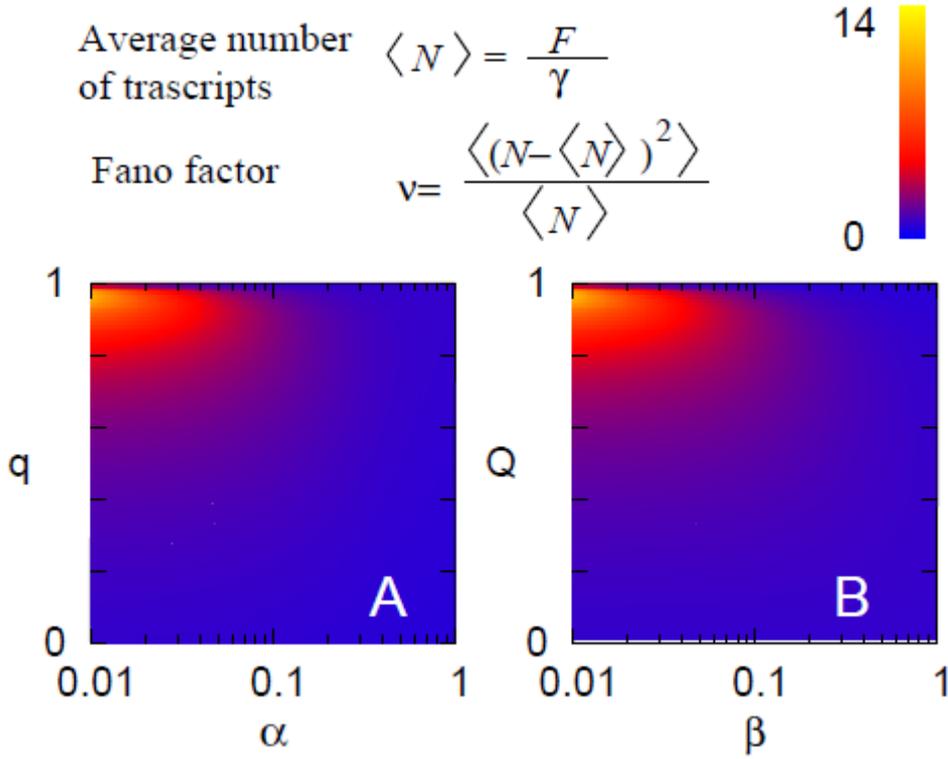

Figure 4. Parameter dependence of Fano factor. (A) Fano factor $v=1+\dfrac{(1-q)(q-\alpha)}{(\alpha+1-q)[\alpha+1-q+\alpha/(\langle N\rangle(1-q-\alpha))]}$ in the recruitment model for $\langle N\rangle=F/\gamma=10$. The horizontal axis shows the aspect ratio $\alpha=O'/E$, and the vertical axis shows the recruitment probability $q$. The fluctuations are larger for smaller $\alpha$, where the formation of the closed complex is the rate-limiting step. (B) Fano factor for the dead-end model,

$$v=1+Q\left[(1-Q)-\dfrac{(\gamma/d+1)\alpha\beta^2}{(\alpha+1)^2}\right]\left[(\beta+1-Q)\left(\beta+1-Q+\beta\dfrac{\gamma}{d}+\dfrac{(1+d/\gamma)(\gamma/d)^2\alpha\beta^2}{(\alpha+1)^2}\right)\right]^{-1}$$

with $\dfrac{\gamma}{d}=\dfrac{Q}{\langle N\rangle(\beta+1-Q)}$, for $\langle N\rangle=10$ and $\alpha=1$. The horizontal axis is $\beta=(\tau_O+\tau_E)/\tau_{dead}$, the ratio of the average time required for successful firing to the average time taken to remove a dead-end complex, and the vertical axis is the probability of successful firing, $Q$. Small $\beta$ and large $Q$ gives large fluctuations, which enables bust like firing through successive normal firings (from large $Q$) and long silent periods until the dead-end complex is removed (from small $\beta$). The detailed calculations are given in the Text S1.

Tables:

Table 1: Relations between model parameters and the average $\langle \Delta n \rangle$, $\langle t_{on} \rangle$, and $\langle t_{off} \rangle$.

Here, the duration $\Delta$ for a RNAP to transcribe one mRNA after it has been fired from the promoter is added, because in the Golding's experiment the mRNA is already visible when it is being made.

| Parameters | $\langle \Delta n \rangle$ | $\langle t_{on} \rangle$ | $\langle t_{off} \rangle$ |
|---|---|---|---|
| $P_{lac/ara}$ | 2.2 | 6 [min] | 37 [min] |
| Recruitment | $1/(1-q)$ | $(\langle \Delta n \rangle - 1)\tau_E + \Delta$ | $\tau_O$ |
| Dead-end | $1/(1-Q)$ | $(\langle \Delta n \rangle - 1)(\tau_O + \tau_E) + \Delta$ | $\tau_{dead}/Q$ |

# The generation of promoter-mediated transcriptional noise in bacteria

Namiko Mitarai, Ian B. Dodd, Michael T. Crooks, and Kim Sneppen

## Supplement

*1. Promoter strength with recruitment*

With transition rates defined in figure 2(a) we here present equations that connect these intrinsic promoter parameters to the overall activity and occupation probabilities of the promoter. With $\eta$ and $\theta$ being respectively the probability that the promoter is occupied by a closed or open complex, the steady state implies:

$$k_b(1-\eta-\theta-E\theta\frac{l}{v}) = (k_u+O)\eta, \tag{5}$$

$$O\eta + qE\theta = E\theta, \tag{6}$$

which can be solved for $\eta$ and $\theta$. Note that eq.(2) takes into account the occlusion of the promoter the elongating RNAP as it leaves the promoter region of length $l$ with velocity $v$. One finds:

$$\theta = \frac{O}{E(1-q)(1+K) + O(1+El/v) + cE(1-q)/k_b}$$
$$= \frac{1}{(1-q)/\alpha + (1+El/v) + E(1-q)/k_b}, \tag{7}$$

where the last equality expresses $\theta$ in terms of the aspect ratio

$$\alpha = \frac{O}{(1+K)E} = O'/E \tag{8}$$

introduced by [8], where $O' = O/(1+K)$. $\alpha \approx 1$ corresponds to class II promoters, whereas very large or small $\alpha$ effectively corresponds to a single limiting step in transcription initiation (class I). Recruitment into the open complex will predict bunched activity for $\alpha << 1$ where formation of the open complex is rate limiting.

The promoter strength activity given by the rate $E\theta$

$$F = \frac{EO}{E(1-q)(1+K) + O(1+E\frac{l}{v}) + OE(1-q)/k_b} \quad (9)$$

For $q \to 1$ the rate is approximated by:

$$F = \frac{E}{(1+El/v)} \quad (10)$$

This reflects transcription that is governed by elongation initiation, which in turn is limited by self occlusion.

For a promoter where self occlusion is insignificant, the occupation probabilities are simplified to:

$$\eta = \frac{1}{1 + K + O/k_b + O/((1-q)E)}, \quad (11)$$

and

$$\theta = \frac{O}{O + (1-q)(K+1)E + (1-q) \cdot E \cdot O/k_b}. \quad (12)$$

In the limit of large $k_b$, this becomes

$$\theta = \frac{\alpha}{\alpha + (1-q)}. \quad (13)$$

Note that all the equations here recover into the one for the standard three-step model if we set the recruitment probability $q = 0$.

## 2. Promoter strength with dead-end complexes

We can also develop an expression for the strength of the promoter when a fraction $1-Q$ of complexes enter into a non-productive state, from which the only escape is through the removal of the stalled RNAP with a rate $d$ (Fig. 2(b)). We denote the other rates, $k_b$, $k_u$, $O$ and $E$ as before, whereas $\Omega$ is the probability that promoter is occluded by a non-productive complex. As before $\eta$ and $\theta$ is respectively the probability that the promoter is occupied by closed and open complex. Steady state equations for occupancy of the promoter are now:

$$k_b(1 - \eta - \theta - QE\theta\frac{l}{v} - \Omega) = (k_u + O)\eta \quad (14)$$

$$O\eta = E\theta \qquad (15)$$

$$(1-Q)E\theta = d\Omega \qquad (16)$$

These are solved to give probabilities for the promoter to be occupied by respectively a productive open complex ($\theta$), a non-productive complex ($\Omega$), or a closed complex ($\eta$):

$$\theta = \frac{O}{(1+K)E + OE/k_b + (1+QEl/v)O + (1-Q)OE/d}$$

$$\Omega = \frac{E}{d}(1-Q)\theta$$

$$\eta = \frac{E}{O}\theta \qquad (17)$$

The promoter strength is then $QE\theta$, or

$$F = \frac{QEO}{(1+K)E + OE/k_b + (1+QEl/v)O + (1-Q)OE/d} \qquad (18)$$

For $Q=1$ this corresponds to the activity of a promoter without any dead-end complex (the term containing $d$ disappears). Also when the rate of removing dead-end complexes is very high, $d \to \infty$, the main effect of their presence is a reduction of the effective firing rate from open complexes from $E$ to the lower value $QE$.

When the self occlusion is negligible and $k_b$ is large compared to other non-equilibrium rates, we get

$$F = \frac{Qd}{\beta + (1-Q)}, \qquad (19)$$

Here, $\beta = (\tau_O + \tau_E)/\tau_{dead} = d(E+O')/EO'$ is the ratio of the time needed for successful firing and the time to remove a dead-end complex.

## 3. Distribution of time intervals between firings

In figure 1bc and figure 3be, we showed the distributions $P(\Delta t)$ of the time interval between firings, $\Delta t$. We present the equations for $P(\Delta t)$ for our model. We consider the case where the $k_b$ and $k_u$ are large enough that the binding and the unbinding of closed complexes are considered to be in equilibrium and the process is dominated by two steps, the formation of the open complex with a rate $1/\tau_O = O/(1+K)$ and the formation of the elongating complex with a rate $1/\tau_E = E$.

The distribution for the standard model are eqs.(1) and (2) in the main text, and the calculation is given in Materials and Methods.

In the case with supercoiling assisted recruitment with probability $q$, the distribution is simply sum of the distribution for unassisted initiation (eq. (1) in the main text) with a weight $(1-q)$ and the distribution for the single supercoiling assisted step from open complex to elongating complex with a weight $q$, which is given by

$$P(\Delta t) = q \; \frac{\exp\left(-\frac{\Delta t}{\tau_E}\right)}{\tau_E}$$

$$+ (1-q) \; \frac{\left[\exp\left(-\frac{\Delta t}{\tau_O}\right) - \exp\left(-\frac{\Delta t}{\tau_E}\right)\right]}{\tau_O - \tau_E}.$$

In the case $\tau_E \ll \tau_O$, this is approximated as

$$P(\Delta t) = \frac{q}{\tau_E} \; \exp\left(-\frac{\Delta t}{\tau_E}\right)$$

$$+ \frac{(1-q)}{\tau_O} \; \exp\left(-\frac{\Delta t}{\tau_O}\right). \tag{20}$$

Similarly, in the case of the dead-end model, the distribution is approximately the sum of the distribution for successful initiation with a weight $Q$ and the distribution of the silenced periods caused by the dead-end complex with a weight $(1-Q)$ and average $\tau_{dead/Q}$, or

$$P(\Delta t) = Q \; \frac{\left[\exp\left(-\frac{\Delta t}{\tau_O}\right) - \exp\left(-\frac{\Delta t}{\tau_E}\right)\right]}{\tau_O - \tau_E}$$

$$+ (1-Q) \; \frac{\exp\left(-\frac{\Delta t}{\tau_{dead/Q}}\right)}{\tau_{dead/Q}}.$$

This estimation does not take into account the time needed to form a dead-end complex, but this effect is negligible when removing the dead-end complex is the rate limiting step, i.e. $\tau_{dead} \gg \tau_O + \tau_E$.

## 4. Distribution of successive firings due to recruitment

The distribution of $\Delta n$ in the supercoiling assisted recruitment model is determined by the probability to have $\Delta n - 1$ successive recruitments under the condition that *the first firing occurred*[1]. Because the probability of recruitment is $q$, the probability to

---

[1]Precisely speaking, this is different from the $\Delta n$ defined in Section 4.2, because the

have $\Delta n$ is given by

$$P_{\Delta n} = q^{\Delta n-1} \cdot (1-q). \tag{21}$$

The average number of events is given by

$$\langle \Delta n \rangle = \sum_{\Delta n=1}^{\infty} \Delta n P_{\Delta n} = (1-q) \sum_{\Delta n=1}^{\infty} \Delta n q^{\Delta n-1} = \frac{1}{1-q}. \tag{22}$$

## *5. On-time distribution and average for the recruitment model*

Here we calculate the probability for the on-time to be $t_{on}$, considering that the on-time is the duration when a transcription event occurs and then possibly supercoiling assisted recruitment into open complexes occurs successively. First, we calculate on period times for the case when the duration to transcribe one message, $\Delta$, is zero, which means that on-time is given by the sum of the intervals between successive transcription events. (Thus events of $\Delta n = 1$ do not contribute.) The duration between events obeys the exponential distribution $\exp(-t/\tau_E)/\tau_E$.

The probability to have $n$ ($\geq 2$) successive events giving $t_{on}$ is proportional to:

$$P_n(t_{on}) = (1-q)q^{n-1} \frac{1}{\tau_E^{n-1}}$$

$$\times \int_{\sum_{i=1}^{n-1} t_i = t_{on}} [\prod_{i=1}^{n-2} dt_i][\prod_{i=1}^{n-1} e^{-t_i/\tau_E}] \tag{23}$$

$$= (1-q)q^{n-1} \frac{1}{\tau_E^{n-1}} e^{-t_{on}/\tau_E}$$

$$\times \int_{\sum_{i=1}^{n-1} t_i = t_{on}} \prod_{i=1}^{n-2} dt_i. \tag{24}$$

The integral means the volume defined by $\sum_{i=1}^{n-1} t_i = t_{on}$ with $t_i > 0$, given by

$$\int_0^{t_{on}} ds_1 \int_0^{s_1} ds_2 \int_0^{s_2} ds_3 \cdots \int_0^{s_{n-3}} ds_{n-2} = \frac{1}{(n-2)!} t_{on}^{n-2}. \tag{25}$$

Finally we get

---

probability that the duration between successive firing with recruitment being longer than the threshold $\tau_c$ is not zero. However, this probability is so small in the present parameter regime that the difference does not matter in practice.

$$P_n(t_{on}) = (1-q)q^{n-1}\frac{1}{\tau_E^{n-1}}\frac{1}{(n-2)!}t_{on}^{n-2}e^{-t_{on}/\tau_E}. \tag{26}$$

Thus, the probability to have on-time $t_{on}$ is

$$P(t_{on}) = (1-q)\delta(t_{on}) + (1-q)\frac{q}{\tau_E}e^{-t_{on}/\tau_E}$$

$$\times \sum_{n=2}^{\infty}\frac{1}{(n-2)!}\left(\frac{qt_{on}}{\tau_E}\right)^{n-2} \tag{27}$$

$$= (1-q)\delta(t_{on}) + (1-q)\frac{q}{\tau_E}$$

$$\times e^{-t_{on}/\tau_E}e^{qt_{on}/\tau_E}, \tag{28}$$

for $0 \leq t_{on} < \infty$ [^2]. Here the $\delta$ function takes into account that a single event is counted with duration $t_{on} = 0$, and with probability $1-q$ (given that we already started with this single event).

Using $q = 1 - 1/\langle\Delta n\rangle$ one for $t_{on} > 0$ get the on-time distribution

$$P(t_{on}) = \frac{\langle\Delta n\rangle - 1}{\langle\Delta n\rangle}\cdot\frac{1}{\langle\Delta n\rangle\tau_E}e^{-t_{on}/(\langle\Delta n\rangle\tau_E)}. \tag{29}$$

The average on-time is given by $\tau_E(\langle\Delta n\rangle - 1)$, which is the duration of the firing from the open complex multiplied by the average number of the successive recruitment events.

### 1. On-time distribution and average with contribution from events with $\Delta n = 1$.

Now we simply assume that one event gives a fixed on-time $\Delta$. This gives an offset of $\Delta$ for the on-time, thus we have

$$P(t_{on}) = (1-q)\delta(t_{on} - \Delta)$$
$$+ \frac{(1-q)q}{\tau_E}e^{-(t_{on}-\Delta)/(\langle\Delta n\rangle\tau_E)}. \tag{30}$$

---

[^2]: Here $\int_0^\infty \delta(t)dt = 1$, not a half.

The average is given by

$$\langle t_{on} \rangle = q(\tau_E/(1-q) + \Delta) + (1-q)\Delta$$
$$= q\tau_E/(1-q) + \Delta$$
$$= (\langle \Delta n \rangle - 1)\tau_E + \Delta. \tag{31}$$

**2. Determination of the parameters.**

In the experiments by Golding *et al.*, they obtained the average number of transcriptions per burst, $\langle \Delta n \rangle = 2.2$, the average on-time $\langle t_{on} \rangle = 6$ [min] and the average off-time $\langle t_{off} \rangle = 37$ [min]. They also mention that the duration of transcribing 1 message is $\Delta = 2.5$ [min].

The value of $q$ is determined from $\langle \Delta n \rangle$ using eq. (22) as $q = 1 - 1/\langle \Delta n \rangle$. The duration from open complex to the elongation complex $\tau_E$ is determined by eq. (31) using the average on-time $\langle t_{on} \rangle$. The average off-time $\langle t_{off} \rangle$. is given by $\tau_E + \tau_O - \Delta$, which fixes the time to form the open complex, $\tau_O$.

It should be noted that the on-time distribution given in (30) has a sharp peak at $\Delta$. In the simulation data, the frequency are calculated using the bin with width 5[min] as in the experiment by Golding *et al.*, which makes this peak low.

*1. The on-time and off-time distribution in the dead-end complex model*

The distribution of $\Delta n$ and the on-time distribution $P(t_{on})$ in the dead-end model (Fig. 2(b)) are given by similar calculations as the recruitment model: During an on-time the RNAPs take the standard 3-step firing pathway, which takes the time $\tau_O + \tau_E$ per firing, and the probability to take this pathway is $Q$. In particular, if the distribution of $\Delta t$ for the full 3-step firing is a single exponential as in the Class 1 case of the standard model, we can simply get $P(t_{on})$ in the dead-end model by replacing the probability of the recruitment $q$ with the probability to take the firing pathway $Q$ and $\tau_E$ with $\tau_O + \tau_E$.

As a result, we get the following distribution for the number of transcripts per on-time

$$P_{\Delta n} = 1 \cdot Q^{\Delta n-1} \cdot (1-Q). \tag{32}$$

The average number of events is

$$\langle \Delta n \rangle = \frac{1}{1-Q}. \tag{33}$$

The distribution of the on-time is given by

$$P(t_{on}) = (1-Q)\delta(t_{on} - \Delta) + $$
$$(1-Q)\frac{Q}{\tau_O + \tau_E} \times e^{-(t_{on}-\Delta)/(\langle \Delta n \rangle(\tau_O + \tau_E))}. \tag{34}$$

To get the distribution of the off-time, we also need a similar calculation, since if several RNAPs end up in a dead-end complex in a row it makes the off-time longer. When we consider the time spent occupied by a dead-end complex as the off-time, the calculation is parallel to the on-time distribution calculation without $\Delta$ in the recruitment model, replacing $q$ with $(1-Q)$ and $\tau_E$ with $\tau_{dead}$. The only difference is that the first dead-end event also gives the off-time $\tau_{dead}$.

As a result, the probability to have dead-end complexes $n$ times in a row is given by

$$p_n = Q(1-Q)^{n-1}, \tag{35}$$

which gives the average number of the dead-end complexes in a row as $1/Q$. The off-time distribution is given by a single exponential distribution

$$P(t_{off}) = \frac{1}{\tau_{dead}/Q} e^{-t_{off}/(\tau_{dead}/Q)}. \tag{36}$$

Note that the off-time measured from the experiment can be slightly longer than this, because the calculation here does not include either the duration from the last firing to the first formation of the dead-end complex or the duration from the moment of removal of the last dead-end complex to the next initiation.

## 2. Calculation of the Fano factor

When mRNAs are degraded with a rate $\gamma$, the number of mRNAs reaches a steady state. We calculate the Fano factor for the number of mRNAs by using the Fokker-Plank (FP) equations for both the recruitment model and the dead-end model. We again ignore the effect of occlusion of promoters by an elongating complex.

1. The recruitment model

In the recruitment model, the promoter can take the following 3 states:
- No RNAP at the promoter,
- Closed complex at the promoter,
- Open complex at the promoter.

When the RNAP starts elongation, the promoter goes from state 3 to state 1, with one more mRNA in the system when no recruitment occurs, while the promoter goes back to the state 3 with one more mRNA in the system when the recruitment occurs.

We define the probability $f_i(n;t)$ with $i=1,2,3$ as the probability to have $n$ mRNAs in the system *and* that the promoter takes the state $i$ at time $t$. The FP equations for $f_i(n;t)$ are given as follows:

$$\frac{df_1(n;t)}{dt} = -k_b f_1(n;t) + k_u f_2(n;t) + (1-q)E f_3(n-1;t) - \gamma\left[n f_1(n;t) - (n+1)f_1(n+1;t)\right],$$

(37)

$$\frac{df_2(n;t)}{dt} = k_b f_1(n;t) - k_u f_2(n;t) - O f_2(n;t) - \gamma\left[n f_2(n;t) - (n+1)f_2(n+1;t)\right], \tag{38}$$

$$\frac{df_3(n;t)}{dt} = O f_2(n;t) + qE f_3(n-1;t) - E f_3(n;t) - \gamma\left[n f_3(n;t) - (n+1)f_3(n+1;t)\right]. \tag{39}$$

The probability is normalized so that $\sum_{i=1}^{3}\sum_{n=0}^{\infty} f_i(n;t) = 1$. The average number of mRNAs $\langle N \rangle$ and variance $\langle \delta N^2 \rangle = \langle (N-\langle N \rangle)^2 \rangle$ are given by

$$\langle N \rangle = \sum_{i=1}^{3}\sum_{n=0}^{\infty} n f_i(n;t), \tag{40}$$

$$\langle \delta N^2 \rangle = \langle N^2 \rangle - \langle N \rangle^2 = \sum_{i=1}^{3}\sum_{n=0}^{\infty} n^2 f_i(n;t) - \langle N \rangle^2, \tag{41}$$

respectively, and the Fano factor is given by $v = \langle \delta N^2 \rangle / \langle N \rangle$.

In order to calculate them using FP equations, we define the generating functions

$$F_i(z;t) = \sum_{n=0}^{\infty} z^n f_i(n;t). \tag{42}$$

The moments are obtained from

$$1 = \sum_i F_i(1;t), \tag{43}$$

$$\langle N \rangle = \sum_i \frac{dF_i(z;t)}{dz}\Big|_{z=1}, \tag{44}$$

$$\langle N(N-1) \rangle = \sum_i \frac{d^2 F_i(z;t)}{dz^2}\Big|_{z=1}. \tag{45}$$

Multiplying eqs.(37)-(39) by $z^n$ and taking summation of $n$ from zero to infinity, we get the equations for the generating function $F_i(z;t)$. Using the derivatives of the equations and the normalization condition (43) in the steady state (i.e. $\frac{d}{dt} f_i(z;t) = 0$), we can calculate the moments. As a result, we get

$$\langle N \rangle = \frac{E}{\gamma} \frac{O'}{O' + (1-q)E(1+O'/k_b)}, \tag{46}$$

$$\frac{\langle \delta N^2 \rangle}{\langle N \rangle} = 1 - \langle N \rangle + \frac{E}{\gamma} \cdot \frac{k_b O + \gamma q(\gamma + k_u + k_b + O)}{(\gamma + k_u + k_b + O)(E + \gamma - qE) + k_b O}. \tag{47}$$

If equilibrium binding and unbinding are fast enough, i.e, $k_b, k_u \gg O, E, \gamma$, we get

$$\langle N \rangle = \frac{E}{\gamma} \frac{O'}{O' + (1-q)E} = \frac{E}{\gamma} \frac{\alpha}{1-q+\alpha}, \tag{48}$$

$$\nu = \frac{\langle \delta N^2 \rangle}{\langle N \rangle} = 1 + \frac{(1-q)(q-\alpha)}{(\alpha+(1-q))[\alpha+(1-q)+\gamma/E]}, \tag{49}$$

where we used the aspect ration $\alpha = O'/E$. Equation (49) is shown in figure 4(a). In case $\alpha \ll q$ and $\alpha \ll 1-q$, we get

$$\langle N \rangle = \frac{E\alpha}{\gamma(1-q)}, \tag{50}$$

$$\nu = \frac{\langle \delta N^2 \rangle}{\langle N \rangle} = 1 + \frac{q}{(1-q) + \gamma/E}, \tag{51}$$

and the recruitment model gives larger fluctuations than a simple Poisson process.

2. The dead-end model

The promoter in the dead-end model can take the following 4 states:
- No RNAP at the promoter,

- Closed complex at the promoter,
- Open complex at the promoter.
- Dead-end complex at the promoter,

We define the probability $f_i(n;t)$ with $i=1,2,3,4$ as the probability to have $n$ mRNAs in the system AND the that promoter takes the state $i$ at time $t$. The FP equations for $f_i(n;t)$ are given as follows:

$$\frac{df_1(n;t)}{dt} = -k_b f_1(n;t) + k_u f_2(n;t) + QEf_3(n-1;t) + df_4(n;t) - \gamma\left[nf_1(n;t) - (n+1)f_1(n+1;t)\right], \tag{52}$$

$$\frac{df_2(n;t)}{dt} = k_b f_1(n;t) - k_u f_2(n;t) - Of_2(n;t) - \gamma\left[nf_2(n;t) - (n+1)f_2(n+1;t)\right], \tag{53}$$

$$\frac{df_3(n;t)}{dt} = Of_2(n;t) - Ef_3(n;t) - \gamma\left[nf_3(n;t) - (n+1)f_3(n+1;t)\right],, \tag{54}$$

$$\frac{df_4(n;t)}{dt} = (1-Q)Ef_3(n;t) - df_4(n;t) - \gamma\left[nf_4(n;t) - (n+1)f_4(n+1;t)\right]. \tag{55}$$

The probability is normalized so that $\sum_{i=1}^{4}\sum_{n=0}^{\infty} f_i(n;t) = 1$.

By using the generating functions

$$F_i(z;t) = \sum_{n=0}^{\infty} z^n f_i(n;t), \tag{56}$$

and the normalization conditions in the steady state, we get

$$\langle N \rangle = \frac{QE}{\gamma} \frac{\alpha}{\alpha + 1 + O'/k_b + \alpha(1-Q)E/d}, \tag{57}$$

$$\nu = \frac{\langle \delta N^2 \rangle}{\langle N \rangle} = 1 - \frac{QE}{\gamma}\left[\frac{O'}{O' + (1+O'/k_b)E + O'E(1-Q)/d}\right.$$

$$\left. - \frac{O'(d+\gamma)}{(E+\gamma)(d+\gamma) + O'(d+\gamma)(1+(O+\gamma)/(E+\gamma)/(Ok_b)) + (1-Q)EO'}\right] \tag{58}$$

If equilibrium binding and unbinding are fast enough, i.e, $k_b, k_u \gg O, E, \gamma$, we get

$$\langle N \rangle = \frac{Qd}{\gamma(\beta + (1-Q))}, \tag{59}$$

$$v = \frac{\langle \delta N^2 \rangle}{\langle N \rangle} = 1 + Q \left[ (1-Q) - \frac{(\gamma/d+1)\alpha\beta^2}{(\alpha+1)^2} \right] \left[ (\beta+1-Q) \left( \beta+1-Q+\beta\frac{\gamma}{d} + \frac{(1+d/\gamma)(\gamma/d)^2 \alpha\beta^2}{(\alpha+1)^2} \right) \right]^{-1}$$

(60)

with $\beta = (\tau_O + \tau_E)/\tau_{dead} = d(1/O' + 1/E) = (d/E)[(\alpha+1)/\alpha]$. Equation (60) is shown in figure 4(b).